# An Economic Analysis of Electricity Consumption and Electric Vehicle Adoption in California from 2010 to 2021

A PAPER
SUBMIITED TO THE COMMITTEE OF STUDIES FROM
THE DEPARTMENT OF ECONOMICS
UNIVERSITY OF NEW MEXICO
BY

Qi, Pufan

SUPERVISED
BY

Andrew L. Goodkind
Xiaoyang Wang

January 2024

# Abstract

This paper examines the influence of electric vehicles (EVs) on total annual electricity consumption across fifty-eight counties in California from 2010 to 2021. Using a log-linear model to analyze the relationship between electricity consumption and EV ownership, alongside a linear-log model with an instrumental variable (IV) approach, the study finds that annual per capita electricity consumption increased by 0.23% for each additional electric vehicle per 10,000 residents over the twelve-year period. The analysis identifies partisanship—measured as the annual percentage of voter registration for the Democratic Party by county—as a robust instrumental variable. Specifically, a 1% increase in Democratic voter registration corresponds to the adoption of approximately two additional EVs per 10,000 population.

# 1. Introduction

It has been more than a century since the extensive employment of gasoline internal combustion engine vehicles (ICEVs), which has brought us a convenient living style. However, global warming has been more aggravated with each passing decade. Therefore, humans must apply modern technology such as batteries to replace conventional fuel engines for attaining the target of reducing greenhouse gases (GHG) emissions such as carbon dioxide ($CO_2$) in the future. From the data of total U.S. greenhouse gas (GHG) emissions by economic sector in 2021 published by U. S. Environmental Protection Agency (EPA), 28% greenhouse gas emissions emanate from transportation, and 25% are from electricity power plants (EPA 2023). Sixty percent of U.S. electricity is from burning fossil fuels, natural gas, and coal in 2021. The percentage of electricity generation from thermal power plants is over 60% for some states such as Florida and Texas in 2021, on the contrary, the percent of electricity generation by renewable energy (including nuclear, hydro, geothermal, solar, wind and biomass and other) in some states such as California and Washington are more than 50% during the same period (U.S. Energy Information Administration (EIA) 2022). It is well known that clean power is vital for the sustainability of electric vehicles (EVs) eventually. Many countries in the world are trying to reduce the proportion of fossil-fuel power and enhance the percentage of clean power. Take USA and China for example, the proportion of thermal power has been decreased from 68.56% in 2012 to 60.35% in 2022 in the U.S. (EIA 2022), and that has also considerably declined from 80.80% in 2010 to 67.9% in 2020 in China (China Power Yearbook 2022). The added part of power is all clean electricity that is generated from hydro, nuclear, wind, solar and others (including geothermal, wood, biomass, and others). There seems to be a promising blueprint for adopting EVs in some areas such as California with a large fraction of clean power in the following years.

How to enhance the proportion of clean electricity (electricity generated by renewable energy) and how to balance economics and environment due to adopting EVs are the two urgent issues at present. For EVs owners, many characteristics of EVs such as price, convenience of charging and all-electric range (distance an EV can travel on a single full battery charge) are the decisive factors to purchase EVs instead of ICEVs. For the ecosystem, increasing the supply of clean power is rather important for both government and business. Hence, the more demand for high performance EVs, the more need for clean power in the future. According to the data of EIA annual power proportion in USA from 2012 to 2022, it is still challenging to achieve a predominant clean electricity market in a short run in the U.S. The percentage of thermal power (including coal, petroleum, natural gas, and other gas) was about 60% in total power during the eleven years although the proportion declined insignificantly.

The history of wide adoption of EVs is only around one decade, but the market for EVs in America has been increasing tremendously since 2010. According to EIA, the proportion of EVs on the road has escalated from 0.04% in 2012 to 0.82% in 2021 in the whole country, in which BEVs (battery electric vehicles) and PHEVs (plug-in hybrid electric vehicles) consist of EVs because the number of FECVs (Fuel Cell Electric Vehicles) is tiny. Correspondingly, the number of EVs has soared from 94,362 to 2,131,246. For the state level, California has always been the largest holder of EVs from 258,200 in 2016 to 1,264,700 in 2022 following by Florida and Texas according to the U.S. Department of Energy (DOE). Also, the percentage of BEVs has decreased from 50.48% to 37.00% and the percentage of PHEVs has also declined from 45.94% to 35.67% during the same period in California.

Although there is a high proportion of electricity generated from fossil energy during the past ten years and the percentage of EVs in total vehicles is still small-scale, it is imperative and meaningful to investigate the impact of EVs on the mix of electricity by fuel type and environment, especially for areas such as California where the largest registration counts in the country emerge in the last seven years. Nevertheless, economic analyses of EVs and the electricity market remain rare and insufficient until now. The surging adoption of EVs has prompted the public to think about the overall cost of use of EVs, which includes cost-benefits analyses for both EVs stakeholders and ecosystem. Hence this research poses the following question: what is the quantitative relationship between total electricity consumption and adoptions of EVs in California during 2010 to 2021?

In the last decade, the field of engineering drew attention to potential air pollution due to adoption of EVs (Huo et al., 2013; Li et al., 2016), and some research has focused on evaluating the net social benefits of EVs accounting for the diversification of power generation (Huo et al., 2015). To better understand these related questions in view of economics, economic and environmental impacts models have been a popular method to estimate primary energy consumption and costs for EVs development (Gopal et al., 2015). Some scholars from Lawrence Berkeley National Laboratory have developed an integrated modelling platform called Renewable and Electric Vehicles (REV) for assessing the benefits, costs, climate, and primary energy impacts of EVs in 2015. Huo et al. (2013) recommends that EVs should be adopted according to the cleanness of regional power mixes through a life-cycle analysis at provincial level in China. Kantumuchu (2023) also points out that the current challenges and limitations of EVs: battery and its environment impact; charging problem; range; affordability and cyber risk, in which the largest challenge is availability of clean power. Xing et al. (2023) analyzes the EVs market in China from 2012 to 2022 comprehensively, which contains structures, manufacturers, regional discrepancies, technology, and infrastructure as well as different subsidies for both manufacturers and consumers and their correspondent effects. One study examines adoption of EVs in Taiwan with an environmental assessment model, which shows that the success of adoption of EVs depends on transition of petrochemical industry and optimization of power grid (Fukushima et al., 2013).

This study focuses on quantitative analysis between electricity market and EVs in California through performing a panel regression analyses with instrumental variable (IV) approach and explores the relationship between EVs and electricity consumption in fifty-eight counties in California from 2010-2021. Drawing from California Energy Commission data, the objective is to figure out whether adoption of EVs has really brought some noteworthy influence on electricity consumption, accounting for the environmental effect due to transportation. Based on the current literature, we choose partisanship as IV which denotes the percentage of voter registration for democratic party by county level in California during 2010 to 2021. California has been the largest Democrats voter state for many years. With a long history of promoting innovations for environmental sustainability, California created the first economy-wide greenhouse gas limit in the nation and the first climate emissions standards for vehicles. California has made a commitment to 100% renewable energy by 2045 (California government 2023). Obviously, this study needs to examine the influence of partisanship in the role of adoption of EVs in California from 2010 to 2021.One paper examines the impact of partisanship on environmental views involving climate change in United States by literature review with a result of political partisanship has the power to affect environmental opinions (Denicola & Subramaniam, 2014). One article by (Sintov et al., 2020) indicates the relationship between partisanship and adoption of EVs. The authors concentrate on the role of political attribute on the adoption desire of EVs. Through an identity framework, the paper uncovers that Democrats are more likely to adopt EVs than Republicans due to stronger EV symbolic attribute perceptions, which can predict adoption resolution with 545 survey samples from Ohio. We will discuss the selection of partisanship as IV further in the section of methodology. Distribution of voter registration for Democrat (%) by county in California in 2021 is shown as Figure 1. The percentage is over 50% for those metropolitans with large population such as San Francisco Bay Area and Greater Los Angeles.

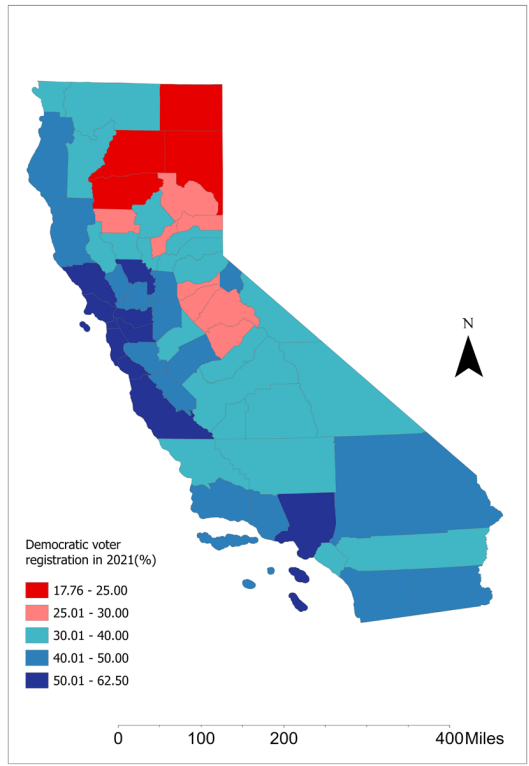

*Figure 1. Distribution of Voter Registration for Democrat (%) by County in California in 2021.*

This paper will proceed as follows: Section two will review the literature on electricity market, EVs in California from 2010 to 2021, also including socioeconomic factors that can potentially exert influence on adoption of EVs such as education, partisanship and homeownership, and empirical strategies that have been widely used for EVs and electricity market in energy economics such as linear or nonlinear regression, simulation method. Section three will contain data and methodology, which will cover data collection, data organization such as data description, data statistics and panel regression analyses with instrumental variable selection and endogeneity test. Section four will show the results and discussion. We will conduct a series of analyses using Stata to examine the quantitative relationship between electricity consumption and EVs, and how those socioeconomic variables impact on it, which is composed of two stages of regressions. Robustness checks are also exhibited in this section. We will also discuss two aspects: EVs consumers in California; solar photovoltaics (PV) electricity and EVs in California. Section five will display limitations and future research and Section six will conclude.

## 2. Literature

This section comprises three aspects in California: electricity market; electric vehicles and factors of impacting on adoption of EVs. We are trying to review the related research about them in the past decade. As one report from Greenhouse Gas Emissions from Electric and Plug-In Hybrid Vehicles by DOE indicated, more EVs implies there will be higher emissions from power plants if the current power sources cannot be transformed. The statistics data about sales of EVs based on EIA shows that EVs and gasoline hybrid contribute to 15.8% of the total annual sales of light-duty vehicles by 2023 in the U.S. and the percents are 12.3% in 2022 and 8.5% in 2021 respectively (EIA 2022). Besides, there are four main benefits brought by solar and home batteries: save electricity bill; protect against outage; control over energy and clean environment. From Figure 2 based on data from EIA, solar energy (PV and thermal) has been the largest renewable energy source in the state since 2016. California state government has made a great mind to accomplish the objective of adopting green technology and substituting NZEVs with ZEVs while contributing to build a low carbon emission electricity market.

*2.1. Electricity Market in California*

Based on the State Profile and Energy Estimates of EIA in April 2023 (EIA 2023), there are some indexes to summarize the current electricity market in California. Renewable power (including hydro power and solar PV power) contributes around half of all electricity generation. Nevertheless, thermal power accounted for 42% and nuclear power served almost 8% of the total net generation. Furthermore, California has the second-largest conventional hydroelectric generation capacity in the United States, next to Washington State, and has been among the top four hydropower markets normally. Although the part from hydropower is steady due to precipitation and snowfall, hydropower cannot be replaced in the short run. Also, the percentage of solar and wind is consistently increasing, and coal fuels power only accounts for a little as shown in Figure 2. The state has decided to terminate all imports of coal-fired generation by 2026. As the largest importer of electricity in the country as well as the largest population, power shortage required the state to import about one-fifth to one-third of its total electricity consumption from neighboring states each year. For the share of the imported electricity in 2021, renewable energy accounted for 31%, hydro shared 16%, nuclear occupied 11% and 23% for unknown sources. What's worse, forest wildfires in California and its neighboring states often plague the safety and stability of both imports and transmission of electricity inside the state. There exist some quandaries of the electricity market in the golden state at present. The first dilemma is that electricity consumption per capita is second lowest last, only more than Hawaii. While total state consumption ranks third after only Texas and Florida in the nation. In addition, the average price of electricity in California ranks second highest and Hawaii has the most expensive price of electricity. Looking at the consumers of electricity in California in 2022, commercial sector shared 46%, 36% for residential sector and approximately 18% was accounted by industrial sector. Only under 0.3% of the total was consumed by the public transportation sector. For promoting EVs in the state and being one part of West Coast Green Highway, California has finished an extensive network of EVs DC fast charging stations placed along interstate 5 (California Energy Commission 2023). Also, the state has demanded all public buses to transit to zero-emission buses step by step before 2029. According to the statistics of EIA, the total system electric generation has increased from 291,046 GWh in 2010 to 302,180 GWh and then decreased to 272,576 GWh. The reason for the declination of electricity generation is probably that California has closed the San Onofre nuclear power plant at the beginning of 2012 and there was a certain amount of population exodus from California during this period (EIA 2023; Wikipedia 2023). The part from both coal-fuel power has dramatically declined while the portion from natural gas is nearly 100,000 GWh, which is approximately 35% of the total electric generation. However, it is promising to see that the electric generation from solar PV has been hugely boosted from 90 GWh in 2010 to 27,179 GWh in 2020. It is the same for the generation from wind, which has increased from 6,172 GWh to 13,708 GWh. From California Energy Commission as shown in Figure 3, annual electricity consumption per capita by county is various by regions in the past decade. Bay area including Alameda County, Marin County, San Francisco County, San Mateo County, Santa Clara County consume more annual electricity per capita than other counties following by greater Los Angeles. Santa Clara ranks first in the total fifty-eight counties and the highest annual amount is more than 20 MWh while the lowest is only around 3 MWh. Considering the influence of EVs alter remarkably providing the configuration of total main power supply and local structure of power grid, the study from (Bellocchi et al., 2019) simulates both similarities and differences on adoption of EVs in Italy and Germany through nonidentical energy schemes. The results specify that the added electricity yields a reduction in $CO_2$ emissions contingent upon renewable power increases notably for Germany. On the contrary, it is advantageous for the Italian power structure from widely adoption of EVs despite at small proportion of clean power.

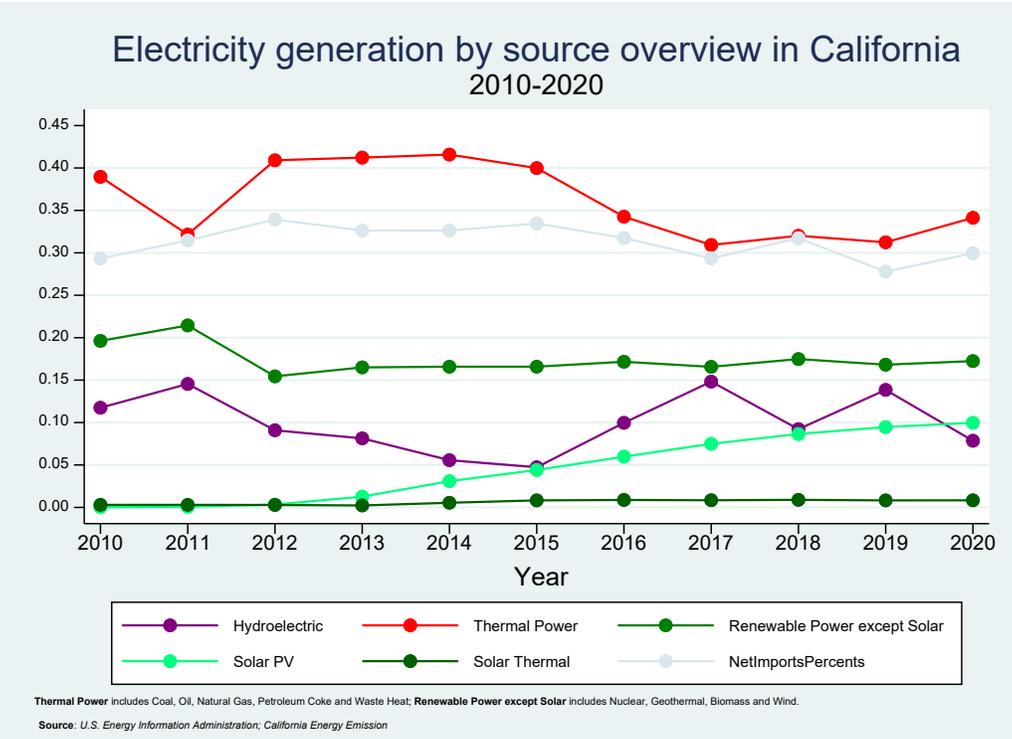

*Figure 2. Overview for Electricity Generation by Source in California from 2010 to 2020.*

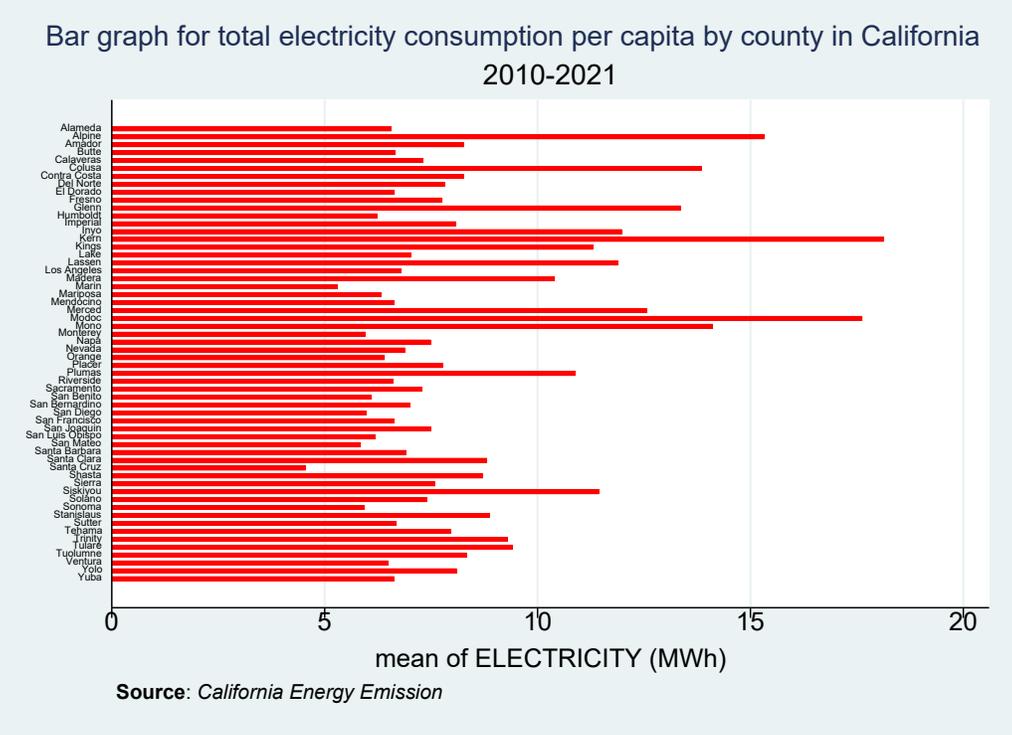

*Figure 3. Bar Graph for total Electricity Consumption per capita by County in California from 2010 to 2021.*

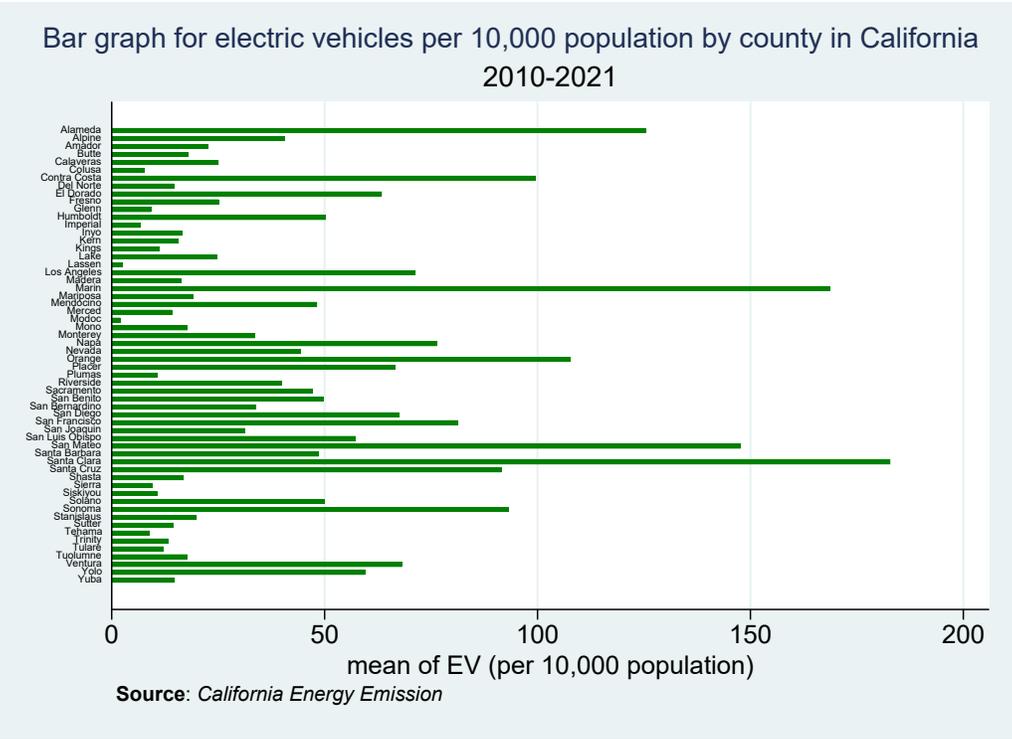

*Figure 4*. Bar Graph for Electric Vehicles per 10,000 population by County in California from 2010 to 2021.

*2.2. Electric Vehicles in California*

The world is promoting the adoption of EVs nowadays. One paper compares both economic and environmental benefits of EVs through a dataset of 1681 multinational vehicles travel information, which presents that the economic benefits and challenges of EVs are commonly identical across regions regardless of variance in fuel prices, driving patterns, and subsidies (He et al., 2019). It's reasonable for Californians have been hugely adopting EVs since 2010 because the state has been a leader of low carbon environment and a state-of-the-art green technology center in the United States for a long time. With plenty of sunshine, large population, expensive gas and, the golden state has pushed out the uncompromising emission standards as well as appreciating the benefits of EVs and plans to target 1.5 million EVs on the road by 2025, and 5 million EVs on the roads by 2030 (California Office of the Governor 2018) (Davis, 2019). According to California government (California Air Resources Board, 2023), California is accelerating a plan of attaining the target of 100% new ZEV sales by 2035, which request that automakers enhance the number of EVs each year since 2026 with new EVs account for from 35% in 2026 to 68% in 2030 and 100% in 2035. As stated by California Air Resources Board (California Air Resources Board, 2023), transportation is the single largest source of emissions and air pollution in California. With increasing penetration of EVs in California, cleaning air and benefiting climate will be feasible. Besides, the California government provides credits to new EVs buyers and tries to increase access to ZEVs for all Californians. As shown in the statistics of light-duty vehicle population in California in 2022 from California Energy Commission (California Energy Commission, 2023), California keeps on with the largest EVs market in the U.S. in 2022 with 40% market share. It also shows that there are 1,111,028 ZEVs with a share of 3.80% in total light-duty vehicles in California by the end of 2022, which includes 763,557 BEVs (2.61%), 335,574 PHEVs (1.15%) and 11,897 FCEVs (0.04%) respectively. Tesla model 3 with population of 266,246 ranks first in terms of cumulative BEVs holding while Tesla model Y (population: 167,955) is the most popular BEV in 2022. Toyota Prius prime is the major PHEVs with cumulative number of 56,525, and Toyota Mirai is the most attractive FCEV with 10,045 holding and it is one of the only six models of FCEVs in California currently. Toyota Prius is regarded as the top one sales of gasoline hybrid vehicle in 2022. Furthermore, as exhibited in Figure 5, the number of BEVs has climbed from 601 in 2010 to 763,557 in 2022, which accounts for 2.61% in all light-duty vehicle population in California in 2022. Similarly, the number PHEVs of has surged from 153 in 2010 to 335,574 in 2022 with a share of 1.15%. For FECVs, the proportion of FCEVs is only 0.04% in 2022 and the holding has also increased from 14 in 2010 to 11,897 in 2022. Additionally, the annual number of BEVs has increased rapidly while both that of PHEVs and FCEVs rose a little. However, the holding of gasoline hybrid vehicles is around two times that of BEVs in 2022.

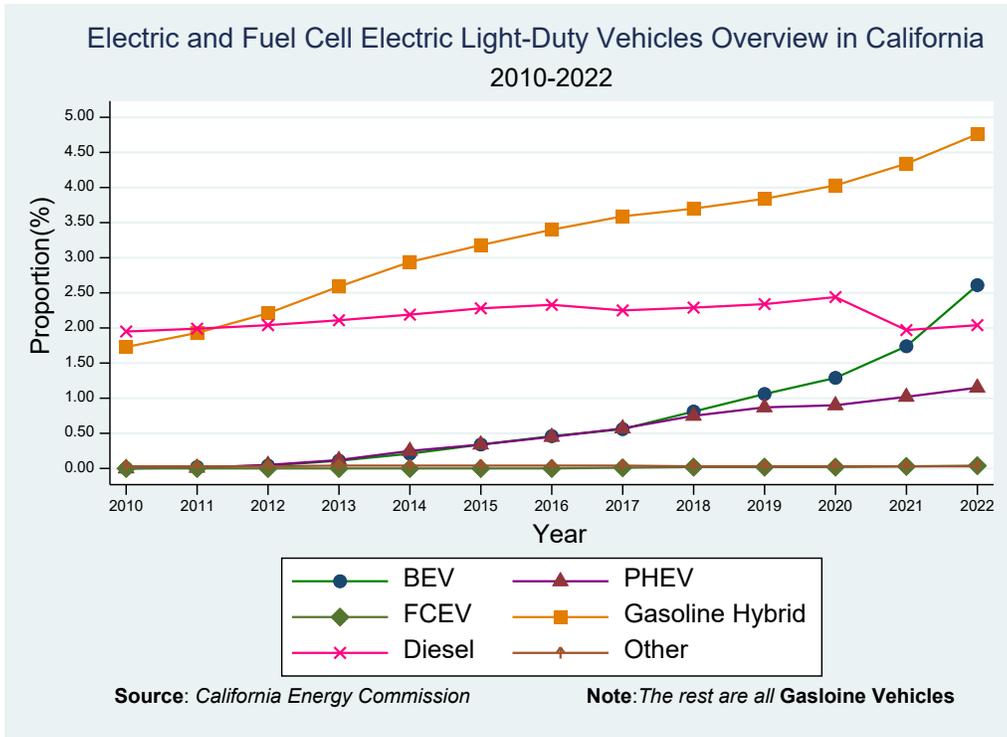

***Figure 5****. Overview for Light-duty Vehicles in California from 2010 to 2022.*

*2.3. Factors of impacting on adoption of EVs*

As Table A2 shows, there are some literatures discussing socioeconomic factors for adoption of EVs. Since the price of EVs is competitive at the nascent period, income is the first factor for any purchasers. Through survey data on the socio-demographic description of 11,037 PEV adopters in California from 2012 to 2017, this study Lee et al. (2019) analyzes four groups with different income level by Bass diffusion models. The results reveal that it is less possible for high income families to adopt PHEVs constantly. Also, PHEVs consumers from high income families account for 49% in total market with only 3.6% share of households in California. However, applying survey data including 26,500 samples, this paper (Christidis P & Focas C, 2019) also investigates the factors affecting the adoption of EVs in the European Union (EU) with machine learning classification model. The conclusion is that the intention of adoption of EVs is significantly associated with income, educational level, and urbanization intensity. Westin et al. (2018) uses logistic regression analyses to explore the relationship between socioeconomic characteristics and adoption of EVs through 1,192 questionaries survey data in Sweden, which shows that both age and education level can promote adopting EVs. Another empirical paper focuses on the impact of homeownership on adoption of EVs with the latest U.S. nationally representative data. The results reveal that homeowners are more willing to adopt EVs than renters. Regardless of income level, the former is three times more likely than renters to purchase an EV with robust statistical significance. Davis W. L. (2019). Considering the current literature and availability of data, this study will employ the following six socioeconomic factors: population density by county; average annual electricity prices; per capita annual personal income; annual college level by county; homeownership by county; COVID-19 pandemic effect, including one geographic factor: average temperature in summer by county. Population density is closely related to urbanization. There usually is more demand for electricity consumption as well as EVs. For example, metropolitans have higher population density where people could commute with EVs. Electricity prices are sensitive to the cost of both electricity consumption and EVs. The higher prices of electricity, the lower desire for using EVs. People with a high educational level are also more likely to adopt new technology. Large houses usually consume more electricity than apartments. Homeownership is a good index to predict consumption of electricity. Since the breakdown of COVID-19, most people stayed at home and many markets were shut, then it is necessary to add the impact of the pandemic on electricity consumption. Also, people travel less during the pandemic, which could reduce the sales of EVs. Higher temperature means more electricity consumption for air conditioners. Therefore, temperature in summer is a good geographic factor.

## 3. Data and Method

This empirical analysis will apply pooled panel regressions which require completeness and accuracy for the annual data with 696 observations. The sample concentrates on all 58 counties in California and the period is from 2010 to 2021, which covers not only the two interesting variables for annual total electricity consumption per capita by county and annual holdings of electric vehicles per 10,000 population by county, but also 8 control variables (annual holdings of gasoline vehicles per 10,000 population; yearly population density by county; average annual electricity prices; per capita annual personal income; yearly college level by county; annual homeownership by county; COVID-19 influence, also including one geographic factor: average annual temperature in summer by county). The descriptions are given in Table 1. The following content will explain them in detail. Besides, we will employ OLS regressions with IV as we mentioned in the introduction. The model and the IV selection will also be explained thoroughly in this section.

*3.1. Data*

The dependent variable is annual total electricity consumption per capita by county, which is the total yearly electricity consumption per capita by county (MWh), including non-residential and residential, and it can be calculated through dividing annual electricity consumption by population. Annual non-residential electricity usage per capita is defined as yearly electricity consumption per capita by county (MWh) by non-residential sector. Annual residential electricity usage per capita is yearly electricity consumption per capita by county (MWh) by residential sector. As EVs can be charged through commercial charging stations and private home charging facilities, it is more efficient and inclusive for total electricity consumption per capita compared to the other two consumptions, this research looks at the total consumption first. All electricity consumption data is from California Energy Commission (California Energy Commission 2023).

Electric vehicles include three types of new energy vehicles that are battery electric vehicles (BEVs), plug-in hybrid electric vehicles (PHEVs) and fuel cell electric vehicles (FCEVs) respectively, which are also called zero-emission vehicles (ZEVs) according to California Energy Commission. Annual holdings of electric vehicles per 10,000 population by county is applied as the focused independent variable in the model. Gasoline vehicles denote annual holdings of gasoline vehicles per 10,000 population that consist of four kinds of light-duty vehicles: gasoline vehicles; gasoline hybrid vehicles; diesel vehicles and other unknown vehicles, which are also called as non- zero emission vehicles (NZEVs). Because the percentage of gasoline vehicles is around 90%, we name all of them as gasoline vehicles for simplication. According to the California Energy Commission, all annual vehicle data is brought up to date in April, which catches the holdings of vehicles on the road during year. Data of vehicles are from California Energy Commission (California Energy Commission 2023).

Population density is the number of populations by county per square miles, which is the yearly data. We calculate the density through dividing the annual population by area. The population data is from the Department of Finance in California (State of California, Department of Finance, December 2021). The data for the area in each county is from Wikipedia (Wikipedia, 2023). Electricity Price is the average annual electricity price, which is the average price of electricity per kWh. Due to availability, we apply the data from Average Price of Electricity & Natural Gas Utilities Los Angeles-Riverside-Orange Counties January, 1979-2022. The data is from Bureau of Labor of Statistics (Bureau of Labor of Statistics, 2023). Per capita annual personal income (unit: $, not seasonally adjusted) is calculated as the personal income of the

residents of a given area divided by the resident population of the area. In computing per capita personal income, BEA uses the Census Bureau's annual midyear population estimates. Personal income is all the income that is received by persons, which includes wages, salaries, supplements to wages and salaries, proprietors' income with inventory valuation and capital consumption adjustments, rental income of persons with capital consumption adjustment, personal dividend income, personal interest income, and personal current transfer receipts, less contributions for government social insurance. The data source is U.S. Bureau of Economic Analysis (U.S. Bureau of Economic Analysis, 2023). Annual college level by county (units: %, not seasonally adjusted) is defined as college level estimate of educational attainment for population 18 years old and over whose highest degree was a bachelor or bachelor above. The data is a multiyear estimation from the American Community Survey (ACS) with 5-year estimation that includes data of a 60-month span. The source is from U.S. Census Bureau (U.S. Census Bureau, 2023). Annual homeownership by county (units: %, not seasonally adjusted) is the homeownership rate of a 5-year estimation by county in California from 2010 to 2021, which is calculated through dividing the estimated total population in owner-occupied units by the estimated total population. A housing unit is defined as owner-occupied on the condition that the owner or co-owner lives in the unit although it is loaned or not 100% paid off. The data source is from Residential Vacancies and Homeownership Annual Statistics of U.S. Census Bureau (U.S. Census Bureau, 2023). The average annual temperature in summer by county (units: degrees Fahrenheit) is the average temperature in summer (from June to August) in each county every year. The data of temperature is from National Oceanic and Atmospheric Administration (NOAA) (National Centers for Environmental information, Climate at a Glance, 2023). COVID-19 is also considered as a control variable and the pandemic really brought plenty of impact to transportation and electricity consumption, which is just a dummy variable in the model. The dummy variable is 0 for the years from 2010 to 2019 and 1 for 2020 and 2021. The annual percentage of voter registration for democratic party by county is an important data to indicate partisanship for this study, which is obtained from the report of registration of voter registration statistics. We select the same period report for each year. The source is from California Secretary of State official website (California Secretary of State, 2023).

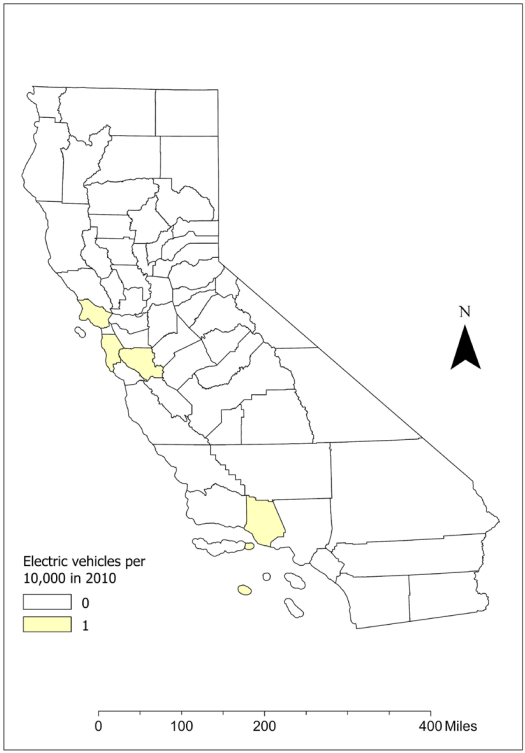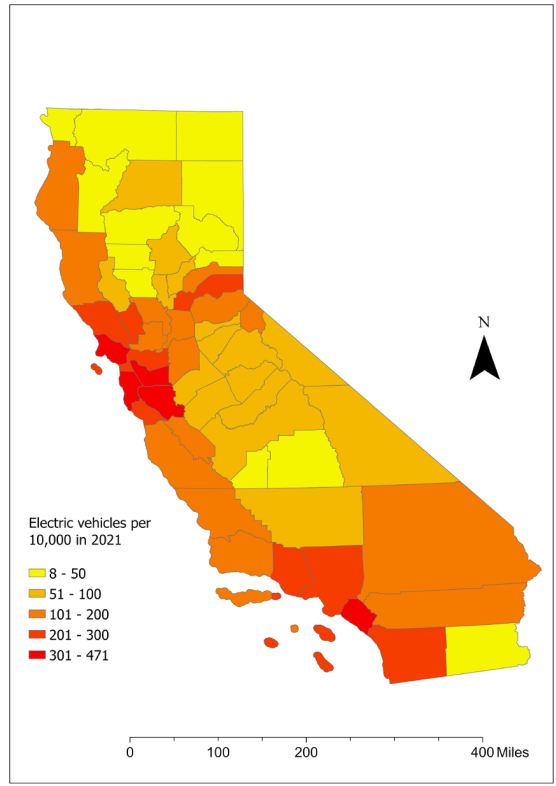

*Figure 6. Distribution of Electric Vehicles per 10,000 Population by County in California in 2010 (left).*
*Figure 7. Distribution of Electric Vehicles per 10,000 Population by county in California in 2021 (right).*

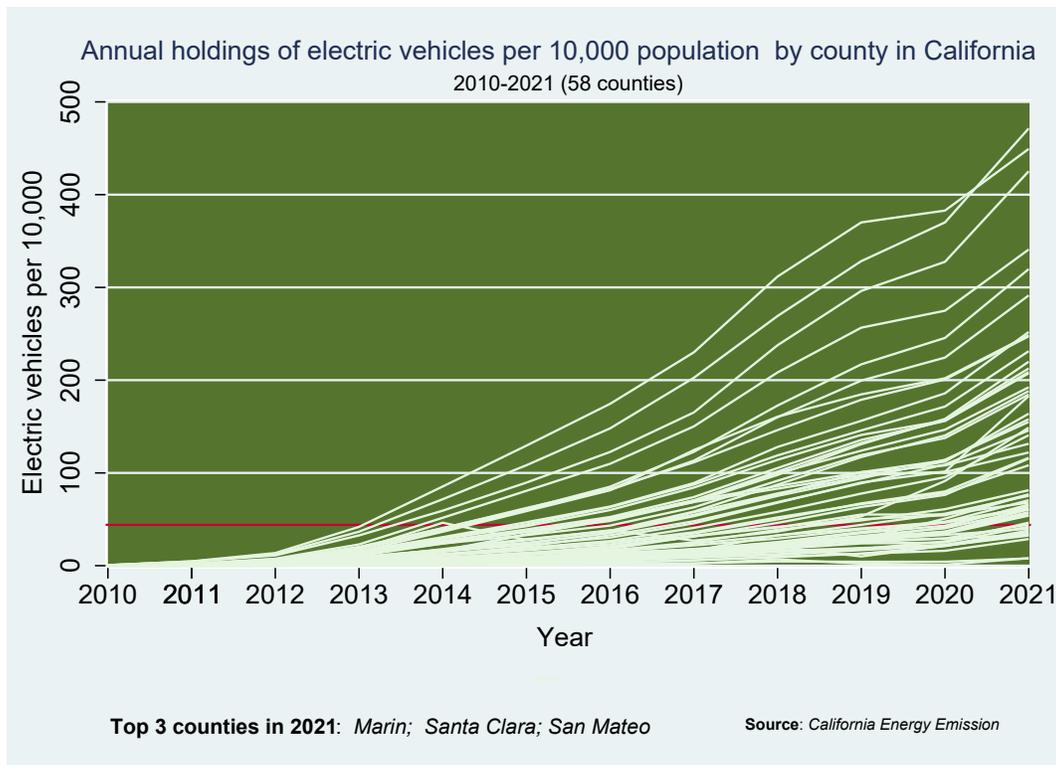

*Figure 8. Electric Vehicles per 10,000 Population by County in California from 2010 to 2021.*

As shown in Figure 6 &7, the geographic distribution of electric vehicles per 10,000 population by county in California in 2010 is just found in San Francisco Bay Area and Great Los Angeles with average around one EV for every 10,000 population. However, the distribution of electric vehicles per 10,000 Population by county in California in 2021 is universe with deep red color. There are around 300 to 400 EVs per 10,000 population in the two metropolitans. Furthermore, based on the statistics of EVs per 10,000 population by county in California from 2010 to 2021, the top three counties of adopting EVs during the 12 years are Marin County, Santa Clara County and San Mateo County respectively, which are all located in the bay area (Figure 8).

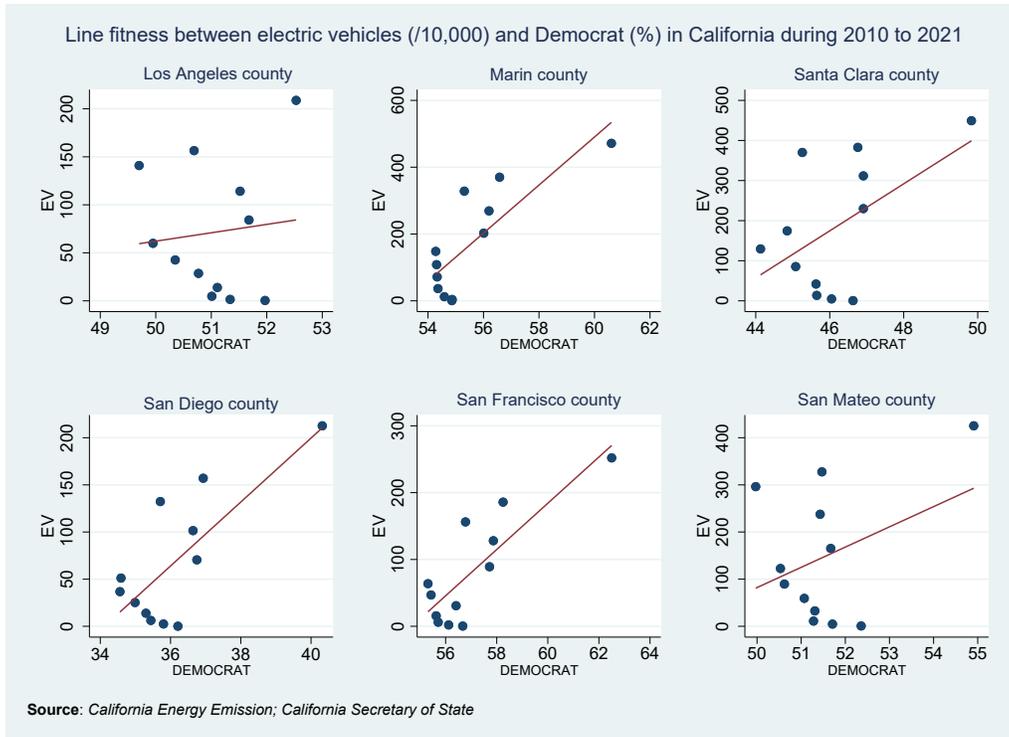

*Figure 9*. Linear Regression Fitting between EV and DEMOCRAT by County in California from 2010 to 2021.

As Figure 9 shown, the two-way linear prediction plots between the holdings of EVs and the annual percentage of voter registration for democratic party by county in California from 2010 to 2021 is obvious (green line from lower left corner to upper right corner), which indicates that more EVs will be adopted if there is a higher percentage of voter registration for democratic party.

*Table 1*. *Description of Variables.*

| Variable | Description | Source |
|---|---|---|
| ELECTRICITY | Total annual electricity consumption per capita by county (including non-residential and residential) by county in California from 2010 to 2021 | California Energy Commission |
| EV | Holdings of ZEVs per 10,000 population, including BEVs, PHEVs and FCEVs by county in California from 2010 to 2021 | California Energy Commission |
| POPDENSITY | Population per square miles by county in California from 2010 to 2021 | Department of Finance in California |
| ELECPRICE | Average price of electricity per kWh by county in California from 2010 to 2021 | Bureau of Labor of Statistics |
| INCOME | Per capita annual personal income (unit: $, not seasonally adjusted) by county in California from 2010 to 2021 | U.S. Bureau of Economic Analysis |
| COLLEGE | Annual college level by county (units: %, not seasonally adjusted) by county in California from 2010 to 2021 | U.S. Census Bureau |
| HOMEOWNERSHIP | Annual homeownership by county (units: %, not seasonally adjusted) (5-year estimate) by county in California from 2010 to 2021 | U.S. Census Bureau |
| TEMP | Average temperature (units: degrees Fahrenheit) from June to August by county in California from 2010 to 2021 | National Centers for Environmental Information (NCEI) |
| COVID | Dummy variable (Covid =1 from 2020-2021) | |
| DEMOCRAT | annual percentage of voter registration for democratic party by county in California from 2010 to 2021 | California Secretary of State |
| GASVEHICLES | Holdings of NZEVs per 10,000 population, including gasoline vehicles, gasoline hybrid vehicles, diesel vehicles and other unknown vehicles by county in California from 2010 to 2021 | California Energy Commission |

Note: The data for the area in each county is from Wikipedia (Wikipedia, 2023).

Table 2 summarizes the statistics of the sample data, which contains 11 variables in total. There are observations following by mean, standard errors, minimum and maximum. The average annual total electricity consumption per capita by county is 8.52 MWh with a range from 4.27 MWh to 20.93 MWh. The average annual holdings of electric vehicles per 10,000 population by county is 44.21 ranging from 0 to 471.40. Similarly, the annual holdings of gasoline vehicles per 10,000 population have a mean of 7,298.68 with a range from 4,279 to 25,190. The average annual percentage of voter registration for democratic party is 38.55% ranging from 17.76% to 62.5%. The annual population density by county has a mean 685.42 per square miles and the minimum is 1.42 and the maximum is 18,756.36. The average annual homeownership by county is 61.65% with a range from 42.58% to 85.69%.

*Table 2*. *Summary Statistics.*

| VARIABLES | N | Mean | Sd | Min | Max |
|---|---|---|---|---|---|
| ELECTRICITY | 696 | 8.52 | 3.01 | 4.27 | 20.93 |
| EV | 696 | 44.21 | 69.66 | 0 | 471.40 |
| TEMP | 696 | 71.96 | 6.45 | 56.27 | 93.60 |
| POPDENSITY | 696 | 685.42 | 2415.70 | 1.42 | 18,756.36 |
| ELECPRICE | 696 | .17 | .03 | .13 | .26 |
| INCOME | 696 | 50,420.62 | 20,807.14 | 26,175 | 164,118 |
| COLLEGE | 696 | 26.15 | 11.12 | 11.70 | 60.20 |
| HOMEOWNERSHIP | 696 | 61.65 | 7.91 | 42.58 | 85.69 |
| DEMOCRAT | 696 | 38.55 | 9.10 | 17.76 | 62.50 |
| COVID | 696 | .17 | .37 | 0 | 1 |
| GASVEHICLES | 696 | 7,298.68 | 1,713.98 | 4,279 | 25,190 |

Note: ELECTRICITY is in MWh; both EV and GASVEHICLES are in per 10,000 population; TEMP is in Fahrenheit; POPDENSITY is in population per square miles; ELECPRICE is in $/kWh; INCOME is in $(current price); COLLEGE is in %; Homeownership is in %; Democrat is in %; COVID is a dummy variable (COVID = 1 when year >=2020). Period: 2010 – 2021.

## 3.2. Panel Regression Analyses

As discussed in the introduction, this study concentrates on an empirical analysis between annual electricity consumption per capita and adoption of EVs, it is common to see that endogeneity occurs when there is a relationship between independent variable and error term in a regression analysis, which may cause bias in the results of regressions. Therefore, one-step regression results cannot be explained as a robust causal effect directly if there are other reasons that lead to a correlation between a treatment variable and an outcome variable. Thus, it is necessary and vital to look at the existence of endogeneity. Because it is impossible to find out all related control variables to build the ordinary least squares (OLS) model. There surely be some omitted control variables, which can cause endogeneity and bias. Also, both measurement errors and simultaneity can make the results worse. For example, there should be some variables that can affect electricity consumption and adoption of EVs at the same time. Therefore, the instrumental variable approach is the best choice for minimizing the errors for the model. There are several reasons for endogeneity existences between electricity consumption and adoption of EVs such as omitted variable and measurement error as well as simultaneity. Many factors can affect electricity consumption. Since we could not find all the control variables to set up the model, it is challenging to try an IV method for convincing results. This empirical analysis is a panel regression combing instrumental variables. Therefore, the key point is that how to show the strong evidence of choosing partisanship as the IV, in detail, why we select the annual percentage of voter registration for democratic party by county as IV for the model? More explanations are provided in the following parts.

*3.2.1. Instrumental Variable Selection*

As literature mentioned above, Democrats are more prone to adopt EVs than Republicans due to stronger EV symbolic attribute perceptions (as shown in Figure 10). The theoretical model for adoption of EVs has described the process in the view of partisanship. There are some symbolic attributes that are related to political identity. Both symbolic attributes and political identity can promote adoption of EVs. At the same time, partisanship can indicate political identity very well. The annual percentage of voter registration for democratic party by county is an ideal index to measure the level of partisanship. Besides, there is not a direct relationship between partisanship and electricity consumption, but partisanship can impact electricity consumption through adoption of EVs or other ways. Democrats are more interested in conservation. There is a much larger relationship between political identity and renewable energy. From the test as shown in Table 5, it indicates that Democrats don't appear to use less electricity. We will do some tests for the potential IV in the next part.

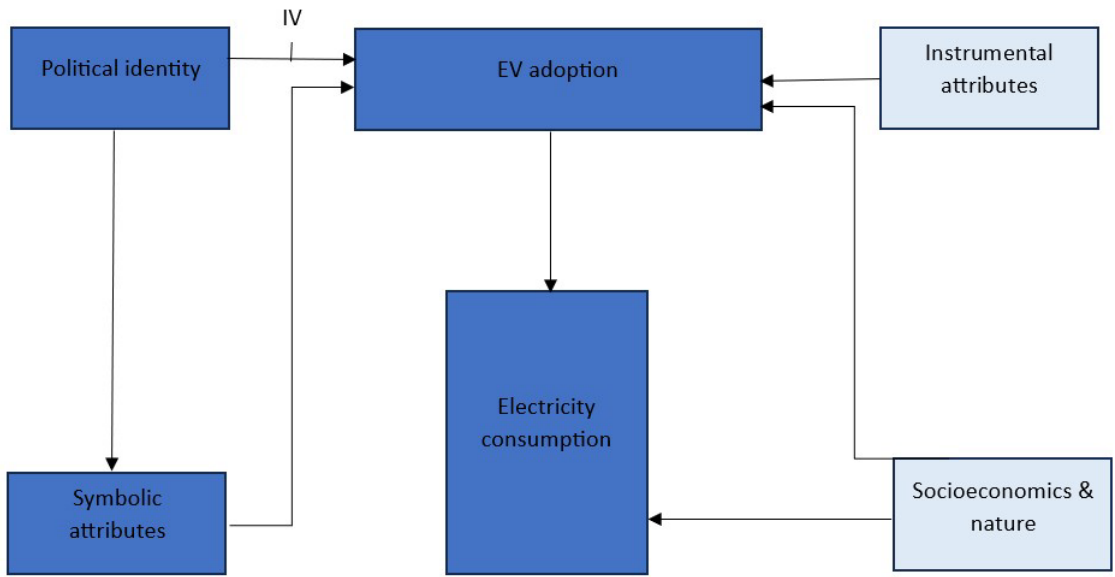

*Figure 10. Conceptual model for adoption of EVs.*

*3.2.2. Endogeneity Test*

It is always hard to find out a legitimate instrument variable. However, it is still worth doing the endogeneity test for a potential IV. In this model, several tests are employed as shown in Table 3. These tests include a test for weak instruments (weak identification test and under identification test), Sargan–Hansen test, Davidson-Mackinnon (1993) test and Hausman-Wu Test. Through the Hausman-Wu test, we can examine whether the error term is correlated to the residuals. Our null hypothesis is that there are no endogenous variables. The IV regression approach uses an instrumental variable correlated with the endogenous predictor variable but not with the error term. Unfortunately, there is no test for exclusion restriction. From the table, it is distinct to see that both results for OLS test and regression with IV are uniform. Instrumental variable is related to the endogenous variable and not related to the interference items. There is not a problem of under identification. In other words, partisanship can be a robust IV, which means that the annual percentage of voter registration for democratic party by county in California can be employed as an IV to evaluate the impact of adoption of EVs on electricity consumption through panel regression.

*Table 3. Summary of Endogeneity Test.*

| Test | Function | Results | Accept or not |
|---|---|---|---|
| Weak identification | Relevant requirement | 3.037 | Yes |
| Under identification | Relevant requirement | 3.068 | Yes |
| Exclusion restriction | Exclusion requirement | Not available | N/A |
| Sargan–Hansen test | Overidentification | 0 | Yes |
| Davidson-Mackinnon Test | Endogeneity | 0.01 | Yes |
| Hausman-Wu Test | Endogeneity | 2.40 | Yes |

According to the literature review and the IV analysis, we propose the following hypothesis:
**Hypothesis**: Annual electricity consumption per capita is positively related to adoption of EVs.

We will apply a log-linear with fixed effects, including robust standard errors to investigate the quantitative relationship between electricity consumption and holdings of EVs. A statistical regression model with instrumental variable that helps predict the most influential exogenous variables. Control variables include yearly population density by county, which can affect electricity consumption; average annual electricity prices also have direct influences on electricity consumption; personal annual income per capita is strongly related to both electricity consumption and adoption of EVs; yearly college level by county can be important role in decision of purchasing EVs; annual homeownership by county can promote electricity consumption; average annual temperature in summer by county is very related to electricity consumption; and annual holdings of gasoline vehicles per 10,000 population. COVID-19 is used as a dummy variable, which has brought a huge impact on both electricity consumption and sales of EVs.

Model between electric vehicles and instrumental variable

$$EV_{ct} = \beta_0 + \beta_1 lnDEMOCRAT_{ct} + \beta_2 X_{ct} + u_c * t + \mu_t + \mu_c + e_{ct} \quad (1)$$

Log-linear model between electricity consumption and electric vehicles

$$lnELECTRICITY_{ct} = \gamma_0 + \gamma_1 \widehat{EV}_{ct} + \gamma_2 X_{ct} + u_c * t + \mu_t + u_c + \varepsilon_{ct} \quad (2)$$

where the dependent variables, $ELECTRICITY_{ct}$ is the annual total electricity consumption per capita by county $c$ at year $t$. where the independent variables, $EV_{ct}$ is the number of annual holdings of electric vehicles per 10,000 population in county c in year t. $X'_{ct}$ are control variables: $TEMP_{ct}$ is average annual temperature in summer in county c in year t. $POPDENSITY_{ct}$ is yearly population density in county c in year t. $ELECTPRICE_{ct}$ is the average annual price of electricity in county c in year t. $INCOME_{ct}$ is personal annual income per capita in county c in year t. $COLLEGE_{ct}$ denotes yearly college level in county c in year t. $HOMEOWNERSHIP_{ct}$ is the annual homeownership rate in county c in year t. $DEMOCRAT_{ct}$ is the IV, which indicates the annual percentage of voter registration for democratic party in county c in year t. $COVID_{ct}$ is a dummy variable, which is 1 when Year is 2020 or 2021. $\mu_t$ are time fixed effects. $\mu_c$ denotes county fixed effects. $u_c * t$ is county time specific trend. $\varepsilon_{ct}$ and $e_{ct}$ are error terms.

## 4. Results and Discussion

*4.1. Results of Model between EV and Instrumental Variable*

From Table 4, we observe the following results: as the annual percentage of voter registration for the Democratic Party by county increases by 1%, the number of electric vehicles (EVs) increases by approximately two per 10,000 population. This relationship is positive and statistically significant, indicating a strong first-stage regression. Additionally, both personal annual income per capita and the yearly college-level education rate by county positively and significantly promote the adoption of EVs. Conversely, annual population density by county exhibits a negative and statistically significant relationship with EV adoption, which is reasonable since more densely populated areas typically have fewer cars. However, this factor is not central to our analysis.

*Table 4. Stage One Regression Results with IV.*

| VARIABLES | (OLS) |
|---|---|
| | EV |
| lnDEMOCRAT | 157.3*** |
| | (37.90) |
| POPDENSITY | -0.0442* |
| | (0.0231) |
| lnELECPRICE | 31.07 |
| | (23.82) |
| INCOME | 0.00209* |
| | (0.00108) |
| COLLEGE | 2.895*** |
| | (0.842) |
| HOMEOWNERSHIP | 0.138 |
| | (0.476) |
| COVID | 18.96 |
| | (26.30) |
| TEMP | -0.472 |
| | (0.634) |
| lnGASVEHICLES | 18.59 |
| | (12.48) |
| County fixed effects | YES |
| Year fixed effects | YES |
| County time trends | YES |
| Constant | -893.3*** |
| | (193.7) |
| Observations | 696 |
| Number of counties | 58 |

Note: Both EV and GASVEHICLES are in per 10,000 population; TEMP is in Fahrenheit; POPDENSITY is in population per square miles; ELECPRICE is in $/kWh; INCOME is in $(current price); COLLEGE is in %; Homeownership is in %; Democrat is in %; COVID is a dummy variable (COVID = 1 when year >=2020). Period: 2010 – 2021. Robust standard errors in parentheses *** $p<0.01$, ** $p<0.05$, * $p<0.1$.

*Table 5. Tests between Outcome Variables and IV.*

| VARIABLES | lnELECTRICITY | lnELECTRICITY |
|---|---|---|
| DEMOCRAT | 0.00437 | 0.00806 |
|  | (0.00614) | (0.00559) |
| POPDENSITY |  | 5.31e-05** |
|  |  | (2.03e-05) |
| lnELECPRICE |  | -0.0520 |
|  |  | (0.0629) |
| INCOME |  | 1.07e-06 |
|  |  | (1.42e-06) |
| COLLEGE |  | -0.000828 |
|  |  | (0.00502) |
| HOMEOWNERSHIP |  | 0.00693*** |
|  |  | (0.00248) |
| lnGASVEHICLES |  | 0.152 |
|  |  | (0.106) |
| TEMP |  | 0.00689** |
|  |  | (0.00292) |
| COVID |  | 0.0370 |
|  |  | (0.0580) |
| County fixed effects | YES | YES |
| Year fixed effects | YES | YES |
| County time trends | YES | YES |
| Constant | 1.906*** | -0.648 |
|  | (0.246) | (1.031) |
| Observations | 696 | 696 |
| R-squared | 0.543 | 0.592 |
| Number of counties | 58 | 58 |

*4.2. Stage Two Regression Results*

As Table 6 shows, there are three columns for the results. Column one is a normal OLS regression without control variables. The coefficient between the logarithmic value of annual total electricity consumption per capita by county and annual holdings of electric vehicles per 10,000 population by county is 0.000239 but not significant, which means that annual total electricity consumption per capita by county will increase 0.02% as one more EV adopted per 10,000 population. Column 2 is the result of a log-linear regression between electricity consumption and EV combing IV method, but there is no control variable neither. The coefficient is 0.00143 without significant level. The results of column 3 are the best of the three, which is also a log-linear regression between electricity consumption and EV combing IV method, and all control variables are added to the model. The results show that annual total electricity consumption per capita by county will increase 0.23% with each additional EV for 10,000 population with 10% significant level. Furthermore, population density and homeownership are both promoting annual electricity consumption per capita by 0.01% and 0.65% respectively, they are both significant at the level of 5%. Temperature in summer also promotes electricity consumption by 0.73% with 5% significant level. However, the coefficient for electric prices is negative but insignificant as well as personal income per capita, which is the same as educational level. The influence of pandemic is positive for electricity consumption without significant level. Gasoline vehicles also promote electricity consumption but insignificantly. The above information indicates that the adoption of EVs really brought some influences on electricity consumption. People with their own houses are more likely to consume more electricity. Electricity consumption is sensitive to temperature in summer. Those areas such as metropolitans with larger population density are likely to have a higher demand for electricity but fewer EVs compared to suburbs and countryside.

*Table 6. Stage Two Regression Results and Regression Results between IV and Control Variables.*

| VARIABLES | (1) _OLS lnELECTRICITY | (2) _2SLS) lnELECTRICITY | (3) _2SLS lnELECTRICITY |
|---|---|---|---|
| EV | 0.000239 | 0.00143 | 0.00226* |
|  | (0.000288) | (0.00122) | (0.00137) |
| POPDENSITY |  |  | 0.000146** |
|  |  |  | (7.42e-05) |
| lnELECPRICE |  |  | -0.119 |
|  |  |  | (0.0885) |
| INCOME |  |  | -3.39e-06 |
|  |  |  | (4.28e-06) |
| COLLEGE |  |  | -0.00765 |
|  |  |  | (0.00709) |
| HOMEOWNERSHIP |  |  | 0.00646** |
|  |  |  | (0.00294) |
| COVID |  |  | 0.0219 |
|  |  |  | (0.0632) |
| TEMP |  |  | 0.00726** |
|  |  |  | (0.00305) |
| lnGASVEHICLES |  |  | 0.107 |
|  |  |  | (0.121) |
| County fixed effects | YES | YES | YES |
| Year fixed effects | YES | YES | YES |
| County time trends | YES | YES | YES |
| Constant | 2.082*** | 2.090*** | 0.232 |
|  | (0.00412) | (0.00918) | (0.974) |
| Observations | 696 | 696 | 696 |
| Number of counties | 58 | 58 | 58 |

Note: ELECTRICITY is in MWh; both EV and GASVEHICLES are in per 10,000 population; TEMP is in Fahrenheit; POPDENSITY is in population per square miles; ELECPRICE is in $/kWh; INCOME is in $(current price); COLLEGE is in %; Homeownership is in %; Democrat is in %; COVID is a dummy variable (COVID = 1 when year >=2020). Period: 2010 – 2021. Robust standard errors in parentheses *** p<0.01, ** p<0.05, * p<0.1.

# 5. Limitations and Future Research

There are several limitations to this study. First, accurately calculating the electricity consumption of electric vehicles (EVs) at the county level is beyond the scope of this research. Second, inflation adjustments for prices, such as electricity rates and personal expenses, were not considered for simplicity. Third, data collection for some variables, including electricity prices, was incomplete. For future research, larger sample sizes should be employed, although data collection remains a primary challenge. Additionally, the instrumental variable (IV) approach used here is only one of many possible methodologies that could be applied in similar analyses. Furthermore, China is expected to become the largest EV market in the future, where more urgent issues need to be addressed—not only in renewable energy but also in conserving traditional energy sources. We intend to continue related research on EVs regardless of how the market evolves.

# 6. Conclusions

This study examines the adoption of electric vehicles (EVs) in California using a log-linear growth model and a linear-log model combined with instrumental variables. The analysis reveals that an increase of one additional EV per 10,000 people between 2010 and 2021 corresponds to a 0.23% rise in annual electricity consumption per capita. Drawing on a literature review and regression analysis, the results identify partisanship as a robust instrumental variable. We use the annual percentage of voter registration for the Democratic Party to represent partisanship. Specifically, for every 1% increase in Democratic voter registration by county, approximately two additional EVs are adopted per 10,000 residents. Socioeconomic factors such as population density and homeownership also contribute to increased electricity consumption, boosting it by 0.01% and 0.65% per capita annually, respectively. Additionally, summer temperatures increase electricity consumption by 0.73%. The findings further indicate that the rise in electricity consumption due to EV adoption parallels the growth in solar power generation in California over the past twelve years. This article aims to contribute to the empirical analysis of EV adoption from the perspective of electricity consumption. Although the significance level between the logarithm of electricity consumption and EV holdings is 10%, this finding encourages further similar analyses. One challenge is that EVs remain in an early stage of adoption, and current economic analyses are limited in predicting whether EVs will fully replace gasoline vehicles in the future. Nonetheless, several implications arise for governments and policymakers: the critical issue is whether EVs can be powered by truly sustainable green energy rather than temporary solutions; technology plays a vital role in the modern energy transition, and the sustainability of EVs will largely depend on advancements in state-of-the-art technology; political ideology also influences EV adoption, a factor that is difficult to explain solely through economic analysis. Various factors, including homeownership, affect EV adoption. It is prudent for developing countries to begin adopting EVs early, despite the challenges of establishing a modern green power grid. Early adoption facilitates addressing societal challenges associated with new technologies. Although widespread EV adoption remains a long-term goal, it is both feasible and promising for humanity to benefit from modern technology while simultaneously enjoying a greener environment. Looking ahead, future research should focus on a representative county in California with more comprehensive and accurate data for regression analysis. Investigating the quantitative relationship between renewable power generation—such as solar energy—and EV adoption should be a priority, as this would provide valuable insights for policymakers.

Gopal, A., Witt, M., Abhyankar, N., Sheppard, C., & Harris, A. (2015). Battery electric vehicles can reduce greenhouse gas emissions and make renewable energy cheaper in India.

He, X., Zhang, S., Wu, Y., Wallington, T. J., Lu, X., Tamor, M. A., McElroy, M. B., Zhang, K. M., Nielsen, C. P., & Hao, J. (2019). Economic and climate benefits of electric vehicles in China, the United States, and Germany. *Environmental Science & Technology*, *53*(18), 11013–11022. https://doi.org/10.1021/acs.est.9b00531

Huo, H., Cai, H., Zhang, Q., Liu, F., & He, K. (2015). Life-cycle assessment of greenhouse gas and air emissions of electric vehicles: a comparison between China and the U.S. *Atmospheric Environment*, *108*, 107–116. https://doi.org/10.1016/j.atmosenv.2015.02.073

Huo, H., Zhang, Q., Liu, F., & He, K. (2013). Climate and environmental effects of electric vehicles versus compressed natural gas vehicles in China: a life-cycle analysis at provincial level. *Environmental Science & Technology*, *47*(3), 1711–1718. https://doi.org/10.1021/es303352x

Kantumuchu, V. C. (2023). Challenges and limitations of electric vehicles. In J. Natarajan, M. B. Kantipudi, C.-H. Yang, & Y. Shiao, *The Future of Road Transportation* (1st ed., pp. 159–176). CRC Press. https://doi.org/10.1201/9781003354901-8

Lee, J. H., Hardman, S. J., & Tal, G. (2019). Who is buying electric vehicles in California? characterising early adopter heterogeneity and forecasting market diffusion. *Energy Research & Social Science*. https://doi.org/10.1016/j.erss.2019.05.011

Li, N., Chen, J.-P., Tsai, I.-C., He, Q., Chi, S.-Y., Lin, Y.-C., & Fu, T.-M. (2016). Potential impacts of electric vehicles on air quality in Taiwan. *Science of The Total Environment*, *566–567*, 919–928. https://doi.org/10.1016/j.scitotenv.2016.05.105

National Centers for Environmental information, Climate at a Glance, 2023. NOAA National Centers for Environmental information, Climate at a Glance: County Mapping, published December 2023, retrieved on January 5, 2024, from https://www.ncei.noaa.gov/access/monitoring/climate-at-a-glance/county/mapping

Nuclear Energy Agency 2023. https://www.oecd-nea.org/jcms/pl_51126/low-carbon-generation-is-becoming-cost-competitive-nea-and-iea-say-in-new-report

Report of registration of voter registration statistics. California Secretary of State, 2023. https://www.sos.ca.gov/elections/voter-registration/voter-registration-statistics

Sintov, N. D., Abou-Ghalioum, V., & White, L. V. (2020). The partisan politics of low-carbon transport: why Democrats are more likely to adopt electric vehicles than Republicans in the United States. *Energy Research & Social Science.* https://doi.org/10.1016/j.erss.2020.101576
31

# Appendices

*Table 7_A1. Abbreviations.*

| Abbreviations and Definitions | |
|---|---|
| EV | Electric vehicles include battery-electric vehicles (BEV) and plug-in hybrid electric vehicles (PHEV). |
| ZEV | Zero Emission Vehicles, includes BEVs, PHEVs, and FCEVs |
| BEV | Battery Electric Vehicle |
| PHEV | Plug-in Hybrid Electric Vehicle |
| PEV | PEVs include battery-electric vehicle (BEV) and plug-in hybrid electric vehicle (PHEV). |
| FCEV | Fuel Cell Electric Vehicle |
| Non-ZEV | Non- Zero Emission Vehicles |
| ICEV | Internal combustion engine vehicles which is also called non-zero emission vehicles (non-ZEV) |
| Solar PV | Solar photovoltaics |
| kWh | Kilowatt hours |
| MWh | Megawatt hours |
| GWh | Gigawatt hours |

*Table 8_A2. Factors Affecting People's Attitudes towards Electricity Consumption and Adoption of Electric Vehicles.*

| Factor categories | Factors | EVs | Source |
|---|---|---|---|
| Socio-demographic factors | Education | Positive | (Christidis P & Focas C, 2019) <br> (Westin et al., 2018) |
| | Income level | Positive | (Christidis P & Focas C, 2019) <br> (Lee et al., 2019) |
| | Homeownership rate | Positive | Davis W. L. (2019) |

## Acknowledgements

I cannot thank my advisers enough. Dr. Goodkind has been my primary adviser, and I have been working with him on this paper since the summer of 2021. Since I had no prior experience writing academic papers, he has devoted considerable effort to guiding my research. He is always patient and dedicated to my studies at UNM. I truly appreciate his professional and profound knowledge of natural resources and environmental economics. Although I have not always submitted assignments punctually, he has never given up on me and continually motivates me. Dr. Wang has been my co-adviser since the spring of 2023, and he has been very kind to me. I have learned a great deal about econometrics from him since I arrived here in the fall of 2019. I am also very grateful to Professor Villa for her exceptional expertise in econometrics. She has done her utmost to assist me with Stata. This article would not have been possible without her enthusiastic support and generous time. I also wish to express my deep gratitude to my colleague and friend in the department, Thaneshwar Paneru; his suggestions were vital for developing the model with instrumental variables in this paper. Without my family's support, I would not have been able to come to the United States for graduate studies or to dedicate myself to writing this paper. In sum, the gratitude I feel at this moment cannot be adequately expressed in a few simple words.